\documentclass{IEEEtran}
\usepackage{graphicx}
\usepackage{hyperref}
\usepackage{cite}
\usepackage[noorphans,font=itshape]{quoting}

\title{Debian in the Research Software Ecosystem: 
\\A Bibliometric Analysis}

\author{
  \IEEEauthorblockN{Joenio Marques da Costa, Christina von Flach} \\
  \IEEEauthorblockA{Institute of Computing,
    Federal University of Bahia \\
    Email: \{joenio.costa,flach\}@ufba.br}
}

\date{\today}

\begin{document}

\maketitle

\begin{abstract}

\textit{Context}: The Debian system has historically participated in academic
works and scientific projects, with well-known examples including NeuroDebian,
Debian Med, Debsources, Debian Science, and Debian GIS, where the scientific
relevance of Debian and its contribution to the \textit{Research Software}
ecosystem are evident.

{\em Objective}: The objective of this study is to investigate the Debian
system through academic publications, with the aim of classifying articles,
mapping research, identifying trends, and finding opportunities.

{\em Method}: The study is based on a bibliometric analysis starting with an
initial search for the term ``Debian'' in the titles, abstracts, or keywords of
academic publications, using the Scopus database. This analysis calculates
metrics of co-citation, co-authorship, and word co-occurrence, and is guided by
a set of research questions and criteria for inclusion and exclusion to conduct
the bibliometric analysis.

{\em Results}: The study includes a set of articles published across various
fields of knowledge, providing a map of the academic publication space about
Debian. The study’s data will be available in a public repository, reporting
demographic and bibliometric trends, including the most cited articles, active
countries, researchers, and popular conferences.

{\em Conclusion}: Results includes a bibliometric and demographic analysis
identified in publications about Debian, shedding light on the intellectual
structure of academic research. The results of the analyses can help
researchers gain an overview of existing trends in publications about Debian
and identify areas that require more attention from the scientific community.

\end{abstract}


\section{Introduction} 

Software is a new scientific pillar in modern science. Alongside theories and
experiments, it serves as proof of concept for the development and evaluation
of new technological artifacts, such as algorithms, methods, systems, tools,
and other computer-based technologies
\cite{hasselbring_multi-dimensional_2025}.
Scientists not only use software to conduct research but are also its primary
developers. Across various domains of knowledge, researchers develop and build
new software systems as part of their investigations. 

The \textit{motto} ``better software, better research''
\cite{goble_better_2014}, and questions about how to develop, maintain, and
distribute research software gained popularity. In this context, open-source
practices may influence science to the extent that open source software is
considered a fundamental requirement for Open
Science~\cite{noauthor_paris_nodate, flach:sbc:2021}.  This begs the question
whether and how open source software and its models, are scientifically
relevant and can positively contribute to a sustainable ecosystem of software
for science.

Debian \cite{wikipedia_contributors_debian_2024} is a software project with
sustainable characteristics. This project develops one of the oldest and most
widely used free operating systems, Debian, established in 1993 and based on
Linux. It is composed exclusively of free software and is known for its
stability and reliability, serving as the foundation for many other projects,
notably Ubuntu, Linux Mint, Tails, Deepin, and Raspberry Pi OS. These
characteristics suggest that the socio-technical structures of the Debian
project and system possess interesting qualities for the research software
ecosystem, in terms of a sustainable model for the collaborative,
decentralized, and open production, distribution, and use of software.
Initiatives for distributed workflow tools in Computational Biology that use
Debian \cite{moller_community-driven_2010}, collaborative infrastructure
projects for Neuroscience research like NeuroDebian \cite{halchenko_open_2012},
or Debian Med for software distribution in Medicine \cite{tille_debian_2011},
put Debian closer to open science research infrastructures~\cite{unesco:2021}.
Understanding how Debian is used within the research software ecosystem in
academic settings is important. However, there are no reviews available on this
subject.

In this study, we use {\em Bibliometric Analysis} \cite{donthu_how_2021}, a
rigorous method that reveals nuances in the evolution of a specific field and
sheds light on emerging areas, to investigate Debian from the perspective of
{\em Research Software}, mapping the presence of Debian in academic
publications. 
This work stands as a first step towards understanding how the research
software ecosystem and studies on Debian are related, shedding light on
opportunities and challenges in this field. 

\section{Related Work}

In \cite{bezroukov_open_1999}, an analogy is drawn between the open-source
development model and academic research, with open source considered a special
case of academic research and software development viewed as a scientific
activity similar to applied theories. Debian and Red Hat are cited as the two
major players among Linux distributions.

In \cite{coleman_three_2005}, Debian is considered technically one of the most
well-equipped distributions, known for its strong commitment to the ethical
principles of free software. It has organizational modes and a series of
micro-practices that allow for cohesion, sustainability, and growth over time
within its networked production space, creating a community that evolves
sustainably while adhering to the principles of free software.

In \cite{abate_adoption_2017}, Debian is considered one of the largest
``software collections'' in history, with experience of developing research
software in academic settings, technology transfer to the free software
community through active participation, and understanding needs rather than
imposing an academic vision, which often does not solve real-world problems.

\section{Research Strategy}

We conduct a quantitative {\em Bibliometric Analysis} \cite{donthu_how_2021} on
a dataset of research papers, using the R software and the Bibliometrix package
\cite{aria2017bibliometrix} for analyzing, synthesizing, and presenting the
results. Due to the exploratory nature of our work, we only collected data from
one bibliographic research database, Scopus \cite{scopus_2025}.

\subsection{Goal and Research Questions}

The goal of this study is to analyze the volume of scientific publications
available on the Scopus database that mention Debian with the purpose of
ranking the output of researchers and institutions, from the point of view of
researchers interested in Debian and research software.

The following research questions were formulated with the aforementioned goal in mind:

\begin{description}[\IEEEsetlabelwidth{------}\IEEEusemathlabelsep]
  \item[RQ 1.] What is the annual number of publications? (publication count by year)
  \item[RQ 2.] Which papers have been cited the most by other papers? (top-cited papers)
  \item[RQ 3.] Who are the most active researchers, measured by number of published papers? (active researchers)
  \item[RQ 4.] Which countries are contributing the most, based on the affiliations of the researchers? (active countries)
  \item[RQ 5.] Which venues (i.e., conferences, journals) are the main targets of papers? (top venues)
  \item[RQ 6.] What are the most relevant terms and concepts in the field? (top-relevant terms)
\end{description}

\subsection{Data Sources and Eligibility Criteria}

Scopus \cite{scopus_2025} is a bibliographic database containing abstracts and
citations for academic journal articles. It covers nearly 21,000 titles from
over 5,000 publishers, of which 20,000 are peer-reviewed journals in the
scientific, technical, medical, and social sciences.

\subsection{Search Strategy and Data Retrieval}

We followed a structured, systematic literature review approach to ensure
thoroughness. The data were retrieved from the Scopus database on June 5, 2025,
using the query string {\em ``TITLE-ABS-KEY (debian)''}. A total of 473 results
were identified and extracted in CSV format for further analysis by the
Bibliometrix software. Table \ref{table-eligibility-criteria} describes the
criteria used for paper selection.

\begin{table}[!t]
\renewcommand{\arraystretch}{1.3}
\caption{Eligibility criteria.}
\label{table-eligibility-criteria}
\centering
\begin{tabular}{c|l|l}
\hline
\bfseries Criteria & \bfseries Name & \bfseries Feature\\
\hline\hline
  C1 & Documents type & All types\\
  C2 & Search field & Title; Abstract; Keywords\\
  C3 & Document accessibility & All (open and non-open access)\\
  C4 & Publisher data range & All years\\
  C5 & Query expression & {\em TITLE-ABS-KEY ( debian )}\\
  C6 & Language & All languages\\
\hline
\end{tabular}
\end{table}

\subsection{Screening Process}

We narrowed down the 473 results to 431 primary studies, excluding 21 articles
with a publication year before 1993 or later than 2024, 13 documents without an
author, and 8 articles that were not in English.

We also examined at the context in which the term ``Debian'' was used in the
article to fix errors in the data. For instance, the paper {\em ``La
perspectiva profesional en la reforma de la atención primaria de salud: una
aproximación cualitativa''} from 1995, was removed from our dataset, because
the Spanish term ``debían'' was incorrectly identified. The relevant excerpt
from the abstract is reproduced below.

\begin{quoting}
``programas de enfermedades prevalentes {\bf \em debían} venir
elaborados de forma''
\end{quoting}

Such results were excluded from the CSV file before running the analysis with
the Bibliometrix software. In this phase, 11 papers were manually removed,
resulting in a final CSV file containing 420 documents. This data was processed
using the Bibliometrix tool, which found 1270 authors, 300 venues (journal or
conference name), a timespan of 2000 to 2024, 1207 author keywords, 47 authors
of single-authored documents, and 9759 references.

\section{Demographic Trends and Bibliometrics}

To answer the research questions, we analyzed the demographic trends and
bibliometrics of the academic literature citing Debian.
In total, the analysis encompassed 420 papers, written in the English language,
published between 2000 and 2024. Among these were 259 conference papers, 135
articles, 11 books, 11 book chapters, and 4 reviews.

\subsection{RQ 1 - Publication count per year}

\begin{figure}[!t]
\centering
\includegraphics[width=3.4in]{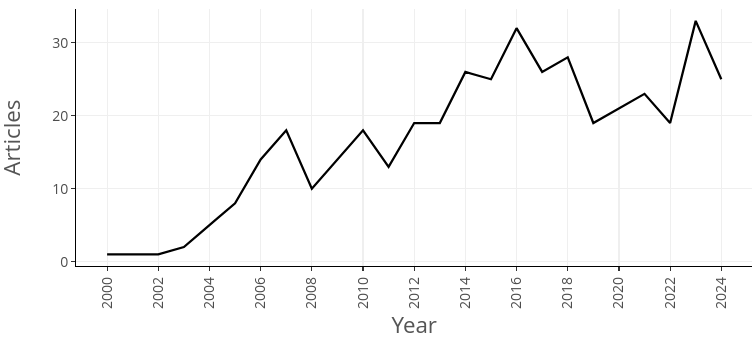}
  \caption{Annual Scientific Production by N. of Documents.}
\label{figure-annual-scientific-production}
\end{figure}

Figure \ref{figure-annual-scientific-production} shows the annual publication
volume of papers mentioning Debian. The first paper indexed by Scopus
mentioning Debian was published in 2000, only 7 years after the Debian release
in 1993.

The first article citing Debian from 2000 is ``Including Diagnostic Information
in Configuration Models'' from the author ``Tommi Syrjänen'', a conference
paper published on the venue ``Computational Logic (CL)''. Debian is mentioned
in the abstract in the following sentence.

\begin{quoting}
``As an example, a subset of the configuration problem for the {\bf \em Debian}
  GNU/Linux system is formalized using the new rule-based language.''
\end{quoting}

The second paper is ``Demudi: The Debian Multimedia Distribution'' by the
author ``Déchelle, F'', published in 2001. The paper introduces the DeMuDi
project, the first multimedia distribution for GNU/Linux.

The third paper is ``Statistically based postprocessing of phylogenetic
analysis by clustering Free'' by the author ``Cara Stockham'', published in
2002. Debian is mentioned in the abstract in the following sentence.

\begin{quoting}
``The Robinson-Foulds distance matrices and the strict consensus trees are
  computed using PAUP (Swofford, 2001) and the Daniel Huson's tree library on
  Intel Pentium workstations running {\bf \em Debian} Linux.''
\end{quoting}

The last article included in our dataset is ``A Comparative Study on
Vulnerabilities, Challenges, and Security Measures in Wireless Network
Security`` published in 2024 by author ``Ahsan Ullah''. Debian is mentioned in
the paper abstract as the environment where the authors generated their
dataset.

\begin{quoting}
``Dataset, generated using Tpot within {\bf \em Debian} operating system,
  serves as the cornerstone of this research, with data collection spanning
  from August 2023 to September 2023, ensuring its timeliness and relevance.''
\end{quoting}

Figure \ref{figure-annual-scientific-production} confirms the trend observed in
other bibliometric studies, indicating an increase in publications over the
years. Between 2000 and 2002, there was one publication per year, followed by
two publications in 2003, five papers in 2004, and eight publications in 2005.
From 2006 onward, the number of publications exceeded ten per year, with peaks
in 2016 (32 papers), 2023 (33 papers), and 2024 (25 papers).

\subsection{RQ 2 - Top-cited papers}

The top two most-cited papers are \textit{``Meep: A flexible free-software
package for electromagnetic simulations by the FDTD method''} (2010), by
\textit{Ardavan F. Oskooi}, with 2,437 citations, and \textit{``SNP-sites:
rapid efficient extraction of SNPs from multi-FASTA alignments Open Access''}
(2016), by \textit{Andrew J. Page}, with 893 citations.  \textit{Ardavan F.
Oskooi}'s paper describes the implementation of an algorithm in a software
named Meep, and Debian is mentioned in the sentence below.

\begin{quoting}
  ``Operating system: Any Unix-like system; developed under {\bf \em Debian}
  GNU/Linux 5.0.2.''
\end{quoting}

\textit{Andrew J. Page}'s article describes the software SNP-sites, and Debian
is mentioned in the sentence below.

\begin{quoting}
  ``It is easy to install through the {\bf \em Debian} and Homebrew package
  managers, and has been successfully tested on more than 20 operating
  systems.''
\end{quoting}

Among the most cited papers, there is a prevalence of articles describing
algorithm implementations and software tools.  According to the
multi-dimensional model for Research Software categorization
\cite{hasselbring_multi-dimensional_2025}, these software tools are research
software of the type ``Modeling, Simulation, and Data Analytics''. All of the
top 10 most cited papers can be understood in the same manner.

\subsection{RQ 3 - Most active researchers}

We followed the approach of other bibliometric/ranking studies
\cite{garousi_systematic_2013} to get an overview of active researchers: we
counted the number of papers published by each author.  The ranking is as
follows: Zacchiroli S. (13 papers), German D.M. (10 papers), Di Cosmo R. (9
papers) and Robles G. (9 papers).

Among the most active researchers, there are publications with studies about
Debian or for Debian. The oldest article by the first author in the ranking,
titled ``The Ultimate Debian Database: Consolidating bazaar metadata for
Quality Assurance and data mining'' (2010), is a study on the metadata of
Debian packages and developers.
The oldest article by the second author in the ranking, titled ``A Model to
Understand the Building and Running Inter-Dependencies of Software'' (2007),
presents a study applied to Debian on package inter-dependencies.

\subsection{RQ 4 - Active countries}

We ranked the most active countries based on the affiliation of the authors who
have published papers mentioning Debian.  The rationale for this ranking is to
know the researchers of which countries (as a group) focus more on using,
contributing or investigating Debian in the context of academic research.

\begin{figure}[!t]
\centering
\includegraphics[width=3.25in]{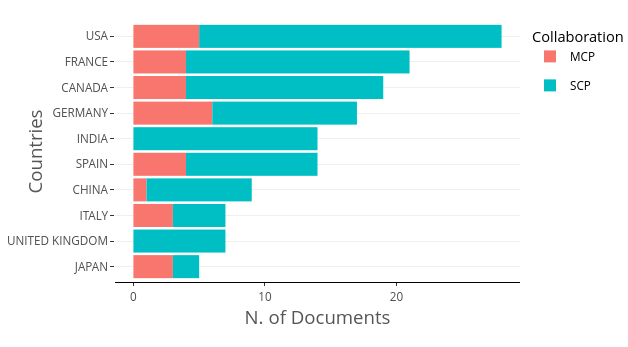}
  \caption{Top 10 Most Productive Countries by N. of Documents based on author affiliations.}
\label{figure-top10-corresponding-author-countries}
\end{figure}

In this analysis, we covered a total of 46 countries. Leading in contributions
are the USA (28 papers), followed by the France (20 papers), Canada (19 papers)
and Germany (17 papers). These findings are presented through a bar chart in
Figure \ref{figure-top10-corresponding-author-countries}, showcasing the
distribution of contributions across the top 10 countries.

\subsection{RQ 5 - Top venues}

To rank the venues, we used the number of papers published in each venue.
Table \ref{table-top10-rank-sources} summarize the top 10 venues. Many major
software engineering conferences and journals are in this list, for example,
the venue with most of the papers (17 papers) is the ``Lecture Notes in
Computer Science (LNCS)''.

\begin{table}[h]
\renewcommand{\arraystretch}{1.3}
\caption{The 10 Sources ranked by number of papers.}
\label{table-top10-rank-sources}
\centering
  \begin{tabular}{p{5cm}|p{1.5cm}|c}
\hline
    \bfseries Name & \bfseries Acronym & \bfseries \# \\
\hline\hline
    Lect. Notes Comput. Science & LNCS & 16 \\
    \hline
    ACM Int. Conf. Proc. Ser. & ICPS & 11 \\
    \hline
    Comput Phys Commun & CPC & 11 \\
    \hline
    Empirical Software Engineering & - & 11 \\
    \hline
    Int Conf Software Engineering & ICSE & 11 \\
    \hline
    IEEE Int. Working Conf. Min Softw. Repos. & MSR & 7 \\
    \hline
    Bioinformatics & - & 6 \\
    \hline
    Adv. Intelligent Sys. Comput. & - & 5 \\
    \hline
    J. Phys. Conf. Ser. & JPCS & 5 \\
    \hline
    ACM Conf Computer Commun Secur & - & 5 \\
\hline
\end{tabular}
\end{table}

\subsection{RQ 6 - Top-relevant terms and frequent words}

\begin{figure}[!t]
\centering
\includegraphics[width=2.8in]{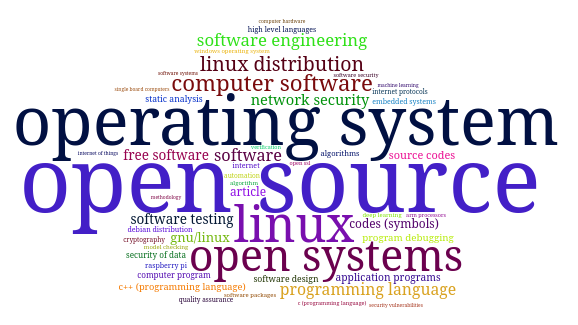}
\caption{Cloud of words from all papers' keywords.}
\label{figure-wordcloud}
\end{figure}

A convenient way to analyze the popular subject areas is to visualize the word
cloud of papers keywords. Figure \ref{figure-wordcloud} shows a word cloud
created with Bibliometrix, using all keywords, including the author keywords
and the Scopus indexed keywords\footnote{The ``author keywords'' are chosen by
the author to best reflect the content of the document, and the ``Scopus
indexed keywords`` are chosen by Scopus and are standardized to vocabularies
derived from thesauri that Elsevier owns or licenses.}.
The terms ``open source'', ``operating system'', ``linux'' and ``open systems''
are among the most popular terms in the set of collected publications. We
notice a lack of relevance for the term ``free software'', despite the
relevance of the Free Software principles that the Debian project adheres.

\section{Discussions}

Below we summarize the main results of each RQ and discuss relevant trends and
findings.

\begin{itemize}
  \item RQ 1. (Publication count by year) 
    \begin{itemize}
      \item The first publication indexed in Scopus citing Debian is from 2000 (7 years after the creation of Debian).
      \item The year with the highest number of papers is 2016 (32 papers) and 2023 (33 papers).
    \end{itemize}

  \item RQ 2. (Top-cited papers) 
    \begin{itemize}
      \item The two top-cited publications have appeared in the years 2010 (2437 citations) and 2016 (893 citations), and are from the fields of Physics and Biology.
      \item These two top-cited publications are software papers contributing with research software, classified as software for ``Modeling, Simulation, and Data Analytics''.
    \end{itemize}

  \item RQ 3. (Active researchers) 
    \begin{itemize}
      \item The four top-active researchers are: Zacchiroli S. (13 papers), German DM. (10 papers), Di Cosmo R. (9 papers) and Robles G. (9 papers).
      \item Among the papers published by the most active researchers we can find studies about Debian and for Debian.
    \end{itemize}

  \item RQ 4. (Active countries) 
    \begin{itemize}
      \item The four most active countries in number of publication mentioning Debian are: USA (28 papers), France (20 papers), Canada (19 papers) and Germany (17 papers).
    \end{itemize}

  \item RQ 5. (Top venues) 
    \begin{itemize}
      \item The best ranked venue with 16 papers is Lecture Notes in Computer Science (LNCS).
      \item Many major software engineering conferences and journals are in this list, such as ICSE and MSR.
    \end{itemize}

  \item RQ 6. (Top-relevant terms) 
    \begin{itemize}
      \item The keyword most frequent among all articles is ``operating system'', followed by ``open source'' and ``linux''.
      \item We notice a lack of relevance for the term ``free software'', despite the relevance of the Free Software principles by the Debian project.
    \end{itemize}
\end{itemize}

\section{Conclusions}\label{conclusions}

The purpose of this study was to draw attention to Debian's prominence in
scholarly publications.  A bibliometric analysis was carried out on the 420
papers published between 2000 and 2024 (24 years) in the Scopus digital
library, guided by six research questions. This bibliometric study revealed the
first impressions about how Debian is mentioned in academic publications,
shedding light on the intellectual structure of academic research citing
Debian.  This study is an initial step toward a more comprehensive bibliometric
analysis of Debian in academic settings and is part of an ongoing research
titled \textit{A theory about Debian as a Sustainable Ecosystem for Research
Software}, as part of the PhD program at the Institute of Computing of the
Federal University of Bahia - Brazil. The study's datasets, analysis plan, and
\LaTeX \ source code are accessible at the following URL:
\url{https://gitlab.com/rslab/debconf2025-academictrack}.

\section*{Acknowledgments}

We would like to thank the Cortext Platform \cite{cortext_manager_v2} team for
the support on how to conduct a bibliometric analysis. Also, would like to
thank Mari Moura for the reviews and the support during the exploratory data
analysis of this study.

\bibliographystyle{IEEEtran}
\bibliography{debconf2025-academictrack}

\end{document}